\documentclass[preprint]{aastex}

\shorttitle{Locally Optimally-Emitting Clouds}
\shortauthors{Korista et al.}
\begin{document}
%
%

\title{Locally Optimally-Emitting Clouds and the Variable Broad
Emission Line Spectrum of NGC 5548}

\author{Kirk T.\ Korista\altaffilmark{1,2}}
\affil{Department of Physics, Western Michigan University}
\affil{Kalamazoo, MI 49008}
\email{korista@wmich.edu}

\and

\author{Michael R.\ Goad\altaffilmark{3}}
\affil{Department of Physics and Astronomy, University of Leicester}
\affil{University Road, Leicester, LE1 7RH, England, UK}
\email{mrg@star.le.ac.uk}




\altaffiltext{1}{Department of Physics \& Astronomy, University of
Kentucky, Lexington, Ky}
\altaffiltext{2}{Visiting Researcher, School of Physics \& Astronomy,
University of St.\ Andrews, Scotland}
\altaffiltext{3}{School of Physics \& Astronomy, University of
St.\ Andrews, Scotland}

\begin{abstract}
In recent work Baldwin et al.\ proposed that in the geometrically
extended broad line regions (BLRs) of quasars and active galactic
nuclei (AGN), a range in line-emitting gas properties (e.g., density,
column density) might exist at each radius, and showed that under these
conditions the broad emission line spectra of these objects may be
dominated by selection effects introduced by the atomic physics and
general radiative transfer within the large pool of line emitting
entities. In this picture, the light we see originates in a vast
amalgam of emitters, but is dominated by those emitters best able to
reprocess the incident continuum into a particular emission line.

We test this ``locally optimally-emitting clouds'' (LOC) model against
the extensive spectroscopic data base of the Seyfert~1, NGC~5548. The
time-averaged, integrated-light UV broad emission line spectrum from
the 1993 {\em HST} monitoring campaign is reproduced via the
optimization of three global geometric parameters: the outer radius,
the index controlling the radial cloud covering fraction of the
continuum source, and the integrated cloud covering fraction. We make
an {\em ad~hoc} selection from the range of successful models, and for
a simple spherical BLR geometry we simulate the emission line light
curves for the 1989 {\em IUE} and 1993 {\em HST} campaigns, using the
respective observed UV continuum light curves as drivers. We find good
agreement between the predicted and observed light curves and lags ---
a demonstration of the LOC picture's viability as a means to
understanding the BLR environment. Finally, we discuss the next step in
developing the LOC picture which involves the marriage of echo-mapping
techniques with spectral simulation grids such as those presented here,
using the constraints provided by a high quality, temporally
well-sampled spectroscopic data set.
\end{abstract}

\keywords{galaxies: active---galaxies: individual (NGC~5548)
galaxies: nuclei---galaxies: Seyfert---line: formation}

\normalsize

\section{INTRODUCTION}

An important goal of quasar research is to understand the origin and
physics of the gas which reprocesses a substantial fraction of the
energy generated by the quasar central engine. Since these broad
line-emitting regions (BLRs) cannot (yet) be imaged directly, we must
infer their properties from the IR -- X-ray spectra.  Spectral
synthesis codes are one of the necessary tools used in the
interpretation of these clues, and sophisticated numerical simulations
of the broad emission lines (BELs) of quasars and active galactic
nuclei (AGN) sprang into existence some 20 years ago (Davidson 1977;
Davidson \& Netzer 1979; Kwan \& Krolik 1981). These were limited to
single-slab photoionization calculations, more because of a lack of
computer power than because of a lack of observational constraints,
though future observations were to better define the breadth of the
BLR.

The multi-wavelength monitoring campaigns of the past decade were
launched to take advantage of the variable nature of the ionizing
continuum and its reverberation signatures in the BELs (Blandford \&
McKee 1982). The results of these campaigns demonstrated the existence
of compact yet geometrically extended and ionization stratified broad
emission line regions whose characteristic sizes scale roughly as
$R_{BLR} \approx 0.1~{\rm pc}~(L_{46})^{1/2}$ (Peterson 1993; Netzer \&
Peterson 1997), where $L_{46}$ is the quasar's mean ionizing luminosity
in units of $10^{46}$ ergs~s$^{-1}$. These and other observations and
the increase in computer power spawned a new generation of
photoionization models (Rees, Netzer, \& Ferland 1989; Goad, O'Brien,
\& Gondhalekar 1993, hereafter GOG93), in which the gas density, column
density, and covering fraction were allowed to vary systematically with
distance from the continuum source in a geometrically thick BLR. In
this scenario each of these parameters are power law functions in
radius, meant to mimic a single pressure law governing the conditions
of the line emitting gas through the BLR. Most recently Kaspi \& Netzer
(1999) applied the pressure law model to their photoionization
calculations and took advantage of the constraints provided by the
observed integrated flux light curves of five emission lines in the
well-studied Seyfert~1, NGC~5548 (Clavel et al.\ 1991, hereafter C91).
Figure~1 shows a plot of its mean UV spectrum from the 1993 {\em HST}
monitoring campaign. Kaspi \& Netzer concluded that the total hydrogen
column densities must be at least $10^{23}$~cm$^{-2}$ at a distance of
1 light-day from the continuum source and that the hydrogen gas
densities here must lie between $10^{11} \la n_H (\rm{cm}^{-3}) \la
10^{12.5}~$.

If the properties of the line emitting gas are controlled by a single
pressure law, what is it that sets and normalizes it to account for the
surprising similarity of quasar/AGN emission line spectra through
several orders of magnitude in luminosity? Can more than one ``pressure
law'' exist? Recently, Baldwin et al.\ (1995) proposed that in a
geometrically extended BLR, a range in line-emitting gas properties
(e.g., density, column density) might exist at each radius, and showed
that under these conditions the BEL spectrum of quasars and AGN may be
dominated by selection effects introduced by the atomic physics and
general radiative transfer within the large pool of line emitting
entities. This model was dubbed ``locally optimally-emitting clouds,''
or LOC. They showed that a typical quasar spectrum results from the
summation of this amalgam of clouds, and that ionization stratification
and the luminosity-radius relationship that produces the similar
spectra are natural outcomes. In the next section we confront the LOC
model with the time-averaged and time-variable spectra of one of the
most intensively studied AGN, NGC~5548, to gain further insights into
the model's strengths and weaknesses and into the physical
characteristics of this object's BLR. In $\S$~3 we discuss the results
and mention a new and potentially powerful technique in deriving the
physical parameters of the BLR (Horne, Korista, \& Goad 1999) --- one
that takes the most general approach to the LOC picture. The
conclusions are given in $\S$~4.

\section{PUTTING A SIMPLE LOC MODEL TO THE TEST}

Here we will confront the predictions of the LOC model with the
observed time-averaged and time-variable BEL spectra of one of the
best-studied Seyfert~1 galaxies, NGC~5548 (C91; Korista et al.\ 1995,
hereafter K95; Peterson et al.\ 1999; references therein). It is not
our intention here to derive the line emitting geometry and dynamics of
NGC~5548, but rather to test the viability of the LOC model under
simple assumptions by comparing the predicted spectrum and emission
line light curves with that of a well-studied AGN.

\subsection{Photoionization Grid Computations \& Assumptions}

Using Ferland's spectral synthesis code, Cloudy (v90.04) (Ferland 1997;
Ferland et al.\ 1998) we generated a grid of photoionization models of
BEL emitting entities, here assumed to be simple slabs (hereafter
``clouds'') each of which we assumed has constant gas density and a
clear view to the source of ionizing photons. The continuum incident
upon the clouds does not include the diffuse emission from other BEL
clouds, nor do we consider the effects of cloud-cloud shadowing. The
grid dimensions spanned 7 orders of magnitude in total hydrogen gas
number density, $7 \le \log~n_H (\rm{cm}^{-3}) \le 14$, and
hydrogen-ionizing photon flux, $17 \le \log~\Phi_H
(\rm{cm^{-2}~s^{-1}}) \le 24$ (see Korista et al.\ 1997, hereafter
K97), and stepped in 0.125 decade intervals in each dimension (3,249
separate Cloudy models). We will call the plane defined by these two
parameters the density -- flux plane. For the present simulations we
assumed all clouds have a single total hydrogen column density, $N_H =
10^{23}~\rm{cm^{-2}}$, though in practice the computations of those
clouds with very low ionization parameter, $U_H \equiv \Phi_H/(n_Hc)
\la 10^{-5}$, stopped when the electron temperature fell below 4000~K.
Below this temperature the gas is mainly neutral and there is very
little contribution to the optical/UV emission lines. Each individual
spectral simulation was iterated until the hydrogen and helium line
optical depths converged to 20\% or better on successive iterations.
The emitted spectrum is not all that sensitive to the cloud column
density over the range $10^{22} \la N_H~(\rm{cm^{-2}}) \la 10^{24}$,
since the emitting volumes of most collisional excitation metal lines
are fully formed within clouds of column densities $10^{22} -
10^{23}~\rm{cm^{-2}}$, given a significant range of gas density and
ionizing flux (K97; Goad \& Koratkar 1998). However, it should be kept
in mind that the variations in an emission line spectrum driven by a
variable ionizing continuum can differ significantly for column
densities spanning $10^{22} - 10^{24}~\rm{cm^{-2}}$ (GOG93; Shields,
Ferland, \& Peterson 1995; Kaspi \& Netzer 1999). We also assumed that
the diffuse emission forms in gas with only thermal motions --- this
may not be the case and local extra-thermal gas motions could have a
significant impact on the diffuse emission spectrum through
de-saturation of optically thick lines, altering the radiative
transfer, and increasing the contribution of photon pumping to the line
emission (Shields, Ferland, \& Peterson 1995). Next, we initially
assumed solar gas abundances (Grevesse \& Anders 1989; Grevesse \&
Noels 1993); however, for reasons discussed below we altered the gas
abundances slightly based upon comparison of the models with the
observed time-averaged spectrum of NGC~5548. Finally, we assumed an
incident continuum spectral energy distribution (SED) that closely
resembles that inferred for NGC~5548 by Walter et al.\ (1994; model A)
using simultaneous {\em IUE} and {\em ROSAT/PSPC} observations (see
also Gondhalekar, Goad, \& O'Brien 1996), and that this continuum is
emitted isotropically and does not change shape substantially as the
luminosity varies. With an average ionizing photon energy of about
84~eV, this continuum is significantly harder than that inferred by
Mathews \& Ferland (1987) for typical quasars. This is necessary in
order to reproduce the heating per photoionization reflected in the
observed mean \ion{C}{4}/Ly$\alpha$ flux ratio ($\sim$~1), large even
for Seyfert~1 spectra.

The equivalent width (EW) contour maps in the density -- flux plane for
six of the seven UV emission line/line blends considered here are shown
in Figure~2a and for \ion{Mg}{2} $\lambda$2800 in Figure~2b. The EW is
proportional to the total energy emitted by the line, and so is {\em a
measure of the continuum reprocessing efficiency for that emission
line}. The value of the EW of each point lying in the grid assumes full
geometric coverage of the continuum source by that particular cloud.
For example, the classical BEL cloud parameters of density and ionizing
flux lie at the location of the ``star'' in the EW contour grids in
Figure~2 (e.g., Davidson \& Netzer 1979). The value of the Ly$\alpha$
EW at this location within the density -- flux plane is approximately
800~\AA\/. It was this predicted EW from a single cloud coupled with
the observed EWs of Ly$\alpha$ that led early researchers to deduce the
value of the cloud covering fraction in quasars ($\sim$~10\%) and
Seyfert galaxies ($\sim$~20\%). The reader may consult K97 for a brief
discussion of the distribution of EW contours for the various emission
lines. All EWs in Figure~2 are measured with respect to the incident
continuum at 1215~\AA\/, and so a ratio of the EW contours of two
emission lines yields their flux ratio. Finally, lines of constant
$\log U_H$ run at 45\degr\/ angles, from lower left to upper right, in
each of density -- flux planes in Figure~2.

\subsection{The UV Broad Emission Line Spectrum of NGC~5548}

In this section we derive from the data a time-averaged, rest frame,
dereddened, velocity-integrated BEL spectrum of NGC~5548, and then
simulate it with a simple LOC model.

\subsubsection{The Time-Averaged Emission Line Spectrum}

Here we establish the velocity-integrated time-averaged ultraviolet BEL
spectrum of NGC~5548. Because of its high quality we chose the
unweighted mean spectrum from the 1993 {\em HST} observing campaign
(Figure~1; see also K95). We use an unweighted mean spectrum because we
want the average emission line fluxes without regard to differences in
S/N in the individual spectra. We list the total measured (observed
frame, reddened) line fluxes in column~2 of Table~1. Since \ion{Mg}{2}
$\lambda$2800 flux was not measured during the 1993 {\em HST} campaign,
we used a slightly smaller value than its mean flux from the 1989 {\em
IUE} campaign to reflect the lower mean continuum flux and its expected
small response to continuum variations. The measured \ion{Mg}{2} flux
is also problematic because of its blending with surrounding
\ion{Fe}{2} emission; see Goad et al.\ (1999) for a recent deblending
analysis for another Seyfert~1. Note that these values are essentially
those that appear in column~5 of Table~24 in K95, with the exceptions
of \ion{C}{4} and \ion{He}{2}~$+$~\ion{O}{3}]. The direct integrated
fluxes of these two sets of emission lines were reported in that table,
and those measurements did not include the region of overlap lying
between them. Here we use their mean fitted fluxes. Goad \& Koratkar
(1998) isolated the UV narrow line spectrum from a single {\em HST}
spectrum (1992 July; Crenshaw, Boggess, \& Wu 1993) when the continuum
and broad emission lines were in a near historic low state (in 1992
April). We list the measured (observed frame, reddened) narrow emission
line fluxes in column~3 of Table~1; most are taken from Goad \&
Koratkar. The \ion{N}{5} value was derived from the analyses in
Korista et al.\ (1995) and that of the \ion{Si}{4}~$+$~\ion{O}{4}]
blend is our recent estimate. The identification of the narrow line
emission contributions of this latter septuplet emission line blend
whose individual narrow line widths are expected to be
$\approx$~5~\AA\/ (FWHM) is difficult since their positions in
wavelength are spread over $\approx$~10~\AA\/. Whatever its value, the
observations would indicate that the narrow line contribution at
$\lambda$1400 is likely to be small.

Correcting the broad line fluxes for reddening was not straight
forward. Galactic \ion{H}{1} measurements place $E(\bv)$ near 0.03
(Murphey et al.\ 1996). However, Kraemer et al.\ (1998) found an
observed {\em narrow} emission line ratio of \ion{He}{2} 1640/4686 that
indicated $E(\bv) \approx 0.07$, placing about $E(\bv) \approx 0.04$
somewhere within NGC~5548. This line ratio is expected to remain near
its simple Case~B value under most conditions, lying near 7 for
conditions in the narrow emission line regions, and thus should be a
robust reddening indicator (Seaton 1978; MacAlpine 1981; Ferland et
al.\ 2000). But does this extra reddening lie within the narrow line
emitting gas, or in a screen covering the narrow and broad emission
line regions plus the continuum, or some combination? For conditions
present within the broad emission line region, this ratio should lie
between 7 and 9 (Ferland et al.\ 2000), and the observed broad-line
ratio might point to the amplitude of the reddening through the
sight-line to the broad emission line region. Unfortunately, the
isolation of the BEL components of the \ion{He}{2} lines is made
difficult due to their breadth and blending with other broad lines
(Wamsteker et al.\ 1990). The broad emission line of $\lambda$1640 is
blended with the extreme red wing of \ion{C}{4} and emission from
\ion{O}{3}] $\lambda$$\lambda$1661,1666. An unreported analysis of K95
attempted to isolate the broad \ion{He}{2} emission using the {\em rms}
spectrum as a guide, and found that approximately 2/3 of the broad
emission from the \ion{He}{2}~$+$~\ion{O}{3}] blend belonged to
\ion{He}{2}, though the significance of this finding is difficult to
quantify. We have adopted this estimated \ion{He}{2}/\ion{O}{3}] ratio
for the present analysis. On the other hand, we know of no attempt to
isolate the BEL component of \ion{He}{2} $\lambda$4686, blended with
moderately strong emission from both \ion{Fe}{2} and the extreme wing
of H$\beta$. Fortunately, whatever the case may be, the reddening
correction is small, and other uncertainties are at least as large.
Here we adopt the Galactic reddening value, $E(\bv) = 0.03$ ($R(V) =
3.1$; extinction curve:  Cardelli, Clayton, \& Mathis 1989), to correct
the BEL and continuum fluxes, and assume the remaining reddening occurs
within the narrow emission line gas of NGC~5548.

The UV BEL fluxes corrected for narrow line fluxes and Galactic
reddening are given in column~4 of Table~1. Finally, column~5 lists the
derived time-averaged UV BEL luminosities ($H_{\rm{o}} =
75~\rm{km~s^{-1}~Mpc^{-1}}$; $q_{\rm{o}} = 0.5$; $z = 0.0172$) and
their adopted uncertainties (in linear luminosity values, not logarithm
values). The latter are rather coarse and meant to merely reflect
estimates of the uncertainties associated with the measurements of
their narrow and broad line fluxes, but do not reflect the
uncertainties associated with the adopted cosmological parameters or
reddening/extinction correction.

Finally, for our choice of SED and measured time-averaged value of
$\log \lambda\/L_{\lambda\/1350} \approx 43.54$ ergs~s$^{-1}$, the
hydrogen-ionizing luminosity is $\log L_{ion} \approx 44.26$
ergs~s$^{-1}$. At this luminosity a $\log \Phi_H = 20$
$\rm{cm^{-2}~s^{-1}}$ in Figure~2 corresponds to a distance from the
continuum source of $R \approx\/~12.6~(75/H_{\rm{o}})$ light-days. It
should be kept in mind, however, that because of reverberation effects,
the measured energy in the UV continuum is not precisely related to the
energy that is measured in the lines, even if the form of the ionizing
SED is known. A monitoring campaign should be of sufficient duration
such that a given line emitting region has been sampled at least once
by the range of Fourier frequency components of the variable incident
continuum. Whether such ``steady-state'' conditions are achievable before
non-reverberation (e.g., dynamical) effects alter the line emitting
regions is uncertain (see Perry, van Groningen, \& Wanders 1994;
Wanders \& Peterson 1996), though the observations do indicate that the
reverberation time scales are generally much shorter than the dynamical
time scales (Peterson 1993).

\subsubsection{Simulating the Time-Averaged UV BEL Spectrum}

The first set of assumptions concerning the integration of emission
from the clouds in our grid involved the simplification of the
geometry: a spherically symmetric distribution of BEL clouds, and we
did not consider either the effects of line beaming (Ferland et
al.\ 1992; O'Brien, Goad, \& Gondhalekar 1994) or continuum beaming
(Wanders \& Goad 1996).

To derive an integrated emission line spectrum, the spectrum of each
cloud lying within the density -- flux plane was assigned a weight in
two dimensions: gas density and distance from the ionizing source
assuming $\Phi_H \propto L/R^2$. Without specifying the particular
shape of the emitting entities, this is equivalent to a two dimensional
($n_H$, $R$), spherically symmetric, function in effective ``cloud''
covering fraction. As in Baldwin et al.\ (1995) and Ferguson et
al.\ (1997) we made the simplifying assumptions that this function is
analytic, separable and a power law in each of the two variables (i.e.,
$f(R) \propto R^{\Gamma}$ and $g(n_H) \propto n_H^{\beta}$; see
equations 1 \& 2 in Ferguson et al.). These assumptions {\em are not
central to the LOC model}, but were chosen merely for their simplicity
given the void of observational constraints. Baldwin (1997) found that
composite quasar spectra were best matched if the power law indices for
the two weighting functions lay near $-1$. This is equivalent to a
cloud covering fraction distribution, $C_f(R,n_H)$, with {\em equal
weighting per decade} in the density -- flux plane. In this case, the
emission line EW (i.e., continuum reprocessing efficiency) contours in
Figure~2 are also proportional to the emission lines' relative
luminosity distributions. Steeper radial and/or flatter gas density
cloud distribution functions concentrate the emission at high continuum
fluxes and gas densities where the emission is mainly thermalized and
inefficient. Flatter radial and/or steeper gas density cloud
distributions concentrate the emission at large radii and low gas
densities. With minimal line thermalization, the line emission from
these latter types of clouds is efficient. However, broad emission line
reverberation and line profile studies indicate that significant line
luminosity must arise from smaller radii as well (Peterson \& Wandel
1999; Wandel, Peterson, \& Malkan 1999). In this analysis we adopted an
index of $-1$ for the weighting along gas density, but allowed for a
range in possible radial covering fraction power law index. The latter
was a parameter in the optimization process, explained below.

Next, using the adopted gas density distribution function we summed the
emission along the density axis for each radius, producing a radial
surface emissivity function for each of the lines and line blends
considered (see Figure~3). We considered densities in the range $8 \le
\log~n_H (\rm{cm^{-3}}) \le 12$ at each radius. We did not include in
our sum the contributions from transparent clouds with very large
$U_H$. While low density clouds lying very near to the continuum source
may have dimensions that rival their distances from the continuum
source, they are also virtually transparent (when Thomson thin) so
their existence is irrelevant for the purposes of this study. Clouds
with gas densities $n_H < 10^8$~cm$^{-3}$ emit unobserved forbidden
lines, and our simple constant density, $10^{23}$~cm$^{-2}$ column
density model clouds become geometrically large relative to the
characteristic size of the BLR (C91; Peterson et al.\ 1991). It is also
true that at the distances from the continuum source at which these low
density clouds are efficient emitters ($\log \Phi_H \la
10^{18}$~$\rm{cm^{-2}~s^{-1}}$), the temperatures of grains, if
present, lie below their sublimation points. Netzer \& Laor (1993)
suggested grain survival as a natural mechanism to cut off the broad
emission at large radii; this would serve to demarcate the boundary
between the BLR and NLR in AGN. Above gas densities of
$10^{12}$~cm$^{-3}$, the majority of clouds are mainly continuum
emitters, and most of the lines are thermalized, assuming thermal local
line widths (Rees, Netzer, \& Ferland 1989; K97). The notable
exceptions to this rule are the excited-state recombination lines of H,
\ion{He}{1}, and \ion{He}{2}, which continue to emit efficiently at
these high densities (see K97). The radiative transfer of the Balmer
lines is probably the least understood and most uncertain of all the
prominent AGN emission lines. In addition the general methods employed
in codes like Cloudy to determine ionization and thermal equilibria
begin breaking down above densities of $10^{12}$~cm$^{-3}$. So while
gas densities of $\sim~10^{14}$~cm$^{-3}$ may be present within the
BLR, and may solve the long standing Ly$\alpha$/H$\beta$ problem
(recently discussed in Netzer et al.\ 1995 and Baldwin 1997), we chose
not to include this gas in our simulations, nor did we use the Balmer
lines to constrain our simulations\footnote{With the gas density
distribution function and upper limit to the gas density of
$10^{12}$~cm$^{-3}$ adopted here, the integrated emission from the
models presented here find Ly$\alpha$/H$\beta$ ratios of about 25.
This is smaller than the classical high-density Case~B ratio of 34, but
still significantly larger than observed, $\sim$~11. Such integrated
ratios are attainable with the gas density distribution adopted here
and an upper limit to the gas density approaching $10^{14}$~cm$^{-3}$
(Baldwin 1997), but we have excluded the extremely dense gas from
consideration here.}. The EWs of all emission lines considered here
peak at or well below gas densities of $10^{12}$~cm$^{-3}$. In summary,
the gas density distribution function was fixed and not part of the
optimization process.

Finally, we fixed the inner radius of the BLR to 1 light-day. This is
not a feature of the LOC picture, and it was done only to accommodate
our chosen analytic cloud covering fraction distribution function of
physical radii, as appropriate for the luminosity of NGC~5548 ($R
\propto (L/\Phi_H)^{1/2}$). As long as this choice of inner radius is
small its impact on the results is minor, since for the adopted
hydrogen-ionizing luminosity the opt-UV emission line gas at smaller
radii must be very high density ($\gg 10^{12}$~cm$^{-3}$) and/or very
high column density. Otherwise the gas will not emit opt-UV emission
lines. In order to account for the response of the emission lines to a
variable ionizing continuum, the emission line surface emissivity
curves were tabulated down to a radius of about \onethird\/ light-day
and out to a radius of about \onethird\/~pc (Figure~3).

To accommodate the fact that several of the measured emission lines are
actually blends, we summed the simulated emission from blended species.
This obviated the problem of relying heavily upon the results of
uncertain deblending analyses. Thus henceforth, Ly$\alpha$ is the sum
of Ly$\alpha$ $\lambda$1216, \ion{He}{2} $\lambda$1216, and O~V]
$\lambda$1218. \ion{Si}{4} is the sum of \ion{Si}{4} $\lambda$1397,
\ion{O}{4}] $\lambda$1402, and \ion{S}{4}] $\lambda$1405. \ion{He}{2}
is the sum of \ion{He}{2} $\lambda$1640, \ion{O}{3}] $\lambda$1663, and
\ion{Al}{2} $\lambda$1670. \ion{C}{3}] is a sum of \ion{C}{3}]
$\lambda$1909, \ion{Si}{3}] $\lambda$1892, and \ion{Al}{3}
$\lambda$1860. \ion{N}{5} $\lambda$1240, \ion{C}{4} $\lambda$1549, and
\ion{Mg}{2} $\lambda$2800 were treated as unblended emission lines.
While \ion{N}{5} is certainly blended with the red wing of Ly$\alpha$,
we assume here that \ion{N}{5} dominates the measurement.

Using the simulated annealing scheme described in Goad \& Koratkar
(1998) to minimize $\chi^2$ between the predicted and observed
emission line luminosities, we varied the outer radius ($R_{out}$), the
power law index on the radial cloud covering fraction ($\Gamma$; see
GOG93), and the normalization to the integrated cloud covering fraction
($C_f$) to fit the time-averaged BEL spectrum in Table~1. The first two
parameters adjust the relative spectrum and the last normalizes the
spectrum to the correct luminosity. For a {\em wide range} of
combinations of these parameters the emission lines belonging to the
$\alpha$-production elements O, Si, and Mg (\ion{O}{6}
$\lambda$1034\footnote{The strength of this line in NGC~5548 has not
been reported in the refereed literature. Based upon reports from other
Seyfert~1 spectra, we have adopted the ratio \ion{O}{6}/\ion{C}{4} =
0.5 as an upper limit to its strength. While we did not include this
upper limit in the optimization process, we did confirm that this upper
limit was realized in all acceptable models for $\Gamma > -1.4$.},
\ion{O}{3}] $\lambda$1663, \ion{Si}{4}~$+$~\ion{O}{4}] $\lambda$1400,
\ion{Si}{3}] $\lambda$1892, \ion{Mg}{2} $\lambda$2800) were all too
strong by factors 1.5--2 compared to their best estimated observed
intensities relative to Ly$\alpha$, \ion{He}{2}, and \ion{C}{4}. Note
that oxygen and silicon are each represented by a resonance line and a
lower ionization intercombination line; each pair of lines together
span large regions in the density -- flux plane (Figure~2a and K97). As
approximate measures of the total heating and photoionization rates,
respectively, the intensities of \ion{C}{4} and Ly$\alpha$--\ion{He}{2}
are to first order independent of the gas abundances. For illustrations
of various emission line sensitivities to gas abundances in the density
-- flux plane, see K97 and Korista, Baldwin, \& Ferland (1998). Given
the results of this preliminary analyses, we considered the possibility
that the gas metal abundances could lie below solar, and we simply
scaled the metal/hydrogen abundance ratios to \onehalf\/ their solar
values. However, we left carbon and nitrogen at their solar abundance
values, since the few spectral constraints, notably \ion{C}{3}] and
\ion{N}{5}, did not indicate subsolar abundances for these two
elements. Carbon, nitrogen, the $\alpha$-elements, and the iron peak
elements all have somewhat different stellar population progenitors and
need not scale together (e.g., Pagel 1997). The He/H abundance ratio
was also left at its solar value. Because the overall metallicity of
the simulated gas is slightly sub-solar, the equilibrium electron
temperatures within the clouds are slightly elevated over their solar
abundance counterparts, and the intensity of the major coolant of the
BEL gas, \ion{C}{4} $\lambda$1549, is enhanced accordingly. This
resulted in a generally closer match to the observed
Ly$\alpha$/\ion{C}{4} ratio for our choice of continuum SED. While we
ascribe no great significance to these adopted gas abundances, they are
less arbitrary than the assumption of solar abundances. A much more
complete analysis of parameter space (cloud distribution functions,
SEDs) will be required to acquire more accurate measures of the gas
abundances.


Figure~4a shows the envelopes in minimum $\chi^2$ (solid curves) as
functions of each of the three parameters, as determined by the
simulated annealing process. The lower dashed lines show the $1\sigma$
confidence level ($\chi^2 = \chi^2_{min} + 1.00 = 2.01$) for one
interesting parameter. The upper dashed lines show the $1\sigma$
confidence contour ($\chi^2 = \chi^2_{min} + 4.72 = 5.73$) for $N - M =
7 - 3 = 4$ interesting parameters. We consider all satisfactory models
to lie below the upper dashed lines and above the solid curves.
Figure~4b shows the confidence contours of $\log \chi^2$ as a function
of $C_f$ and $\Gamma$ for fixed values of $\log R_{out}$ incremented at
0.2 dex. The contours are in steps of 0.25 dex with the outer value
equal to 1.50 dex in every case. Satisfactory models lie within the
bold dashed contour representing $\chi^2 = \chi^2_{min} + 4.72 = 5.73$.
These appear for outer radii $\log R_{out} \ga 1.75$. The relative
emission line spectrum is driven by $\Gamma$ and $R_{out}$, and it is
apparent from the location of the bold contour in Figure~4b that a
broad inverse relationship exists between these two parameters.  Radial
cloud covering fractions which fall off more steeply with radius
generally require larger outer radii. This is because some of the
emission lines are emissive almost exclusively at larger radii (e.g.,
\ion{C}{3}] and \ion{Mg}{2}). Ly$\alpha$ and \ion{C}{4} are emissive at
intermediate and large radii, while \ion{N}{5}, \ion{He}{2}, and
\ion{Si}{4} are also emissive at small radii (see Figures~2 and 3).
Since the line emission from clouds at small radii is inefficient for
the two strongest and best measured emission lines (Ly$\alpha$ and
\ion{C}{4}), larger integrated cloud covering fractions must result
from steeper radial covering fraction distributions. The luminosities
of Ly$\alpha$ and \ion{C}{4} and their ratio provide the tightest
constraints on the models. Together, Figures~4a and 4b show that
satisfactory fits to the time-averaged emission line spectrum are
possible for $R_{out} \ga 60$ light-days, $-1.6 \la \Gamma \la -0.5$,
and $0.33 \la C_f \la 0.80$, although the condition that $C_f \la 50\%$
(as well as \ion{O}{6}/\ion{C}{4}~$\leq 0.5$) constrains $\Gamma \ga
-1.4$. These results are not surprising considering the observed
intensity of \ion{Mg}{2} and its theoretical EW contours in the density
-- flux plane, the analysis of Baldwin (1997), and the observed large
EW of broad Ly$\alpha$ ($\approx 160$~\AA\/), respectively. The models
of Ferland et al.\ (1992), Shields \& Ferland (1993), and Goad \&
Koratkar (1998) used a {\em single cloud} in their attempts to
reproduce the observed properties of Ly$\alpha$ and \ion{C}{4} in
NGC~5548, and their required covering fractions exceeded 30\% -- 40\%.
Thus any model that includes additional emission contributions from
other clouds for other emission lines must necessarily have a larger
covering fraction. Figure~2 also shows that the lowest ionization
parameter clouds included in our models ($-5.5 \la \log U(H) \la -4$)
emit little else but \ion{Mg}{2} $\lambda$2800 (plus optical H,
\ion{He}{1} lines not modeled here; K97).

In choosing one particular fit to the time-averaged spectrum in order
to illustrate that model's emission line variability properties, we
considered the steepest radial covering fraction index for which an
integrated cloud covering fraction $C_f \la 50\%$ resulted: $\Gamma
\approx -1.2$. A steep radial cloud covering fraction distribution
results in broader distributions in the emission line lags (GOG93), and
a broad distribution in lags is observed for NGC~5548. This choice of
$\Gamma$ is also in general agreement with that found by Kaspi \&
Netzer (1999), $-1.33$. Figure~4b shows that good solutions with
$\Gamma < -1.2$ exist, but at the cost of increasingly larger outer
radii and integrated cloud covering fractions. Since we do not know the
origin of the emitting gas, the outer radial boundary of the BLR is
only loosely constrained from any time-averaged, profile-integrated
emission line spectrum. However, a significant contribution of emission
from very large radii will dampen the emission line variability
amplitudes and in most dynamical models will produce narrow emission
lines. Models with integrated cloud covering fractions $C_f > 50\%$ are
surely affected significantly by cloud-cloud shadowing and diffuse
emission from other clouds, and while one could argue that an
integrated cloud covering fraction of even 50\% is significant in this
sense, we adopted $C_f \la 50\%$ as a reasonable validity limit of
these integrated cloud models. These were the only ``filters'' we
applied as we considered our choice of successful model fit to the
time-averaged spectrum.

The first three columns of Table~2 lists the emission lines or line
blends, their simulated time-averaged luminosities, and their simulated
luminosity-weighted radii in light days for the model parameters of
$R_{out} \approx 140$ light-days, $\Gamma \approx -1.2$, and an
integrated cloud covering fraction $C_f \approx 50\%$ ($\chi^2 \approx
3.33$). If grains were present, their temperatures would lie near or
above their sublimation values near this choice of outer radius. All
predicted time-averaged line luminosities are within 0.1~dex of their
target values (Table~1), based upon the mean spectrum of the 1993 {\em
HST} campaign. This and similar models predict the following line
ratios within the blends:  (\ion{He}{2} $\lambda$1216 $+$ O~V]
$\lambda$1218)/Ly$\alpha$ $\lambda$1216 $\approx 0.06$, \ion{O}{4}]
$\lambda$1402/\ion{Si}{4} $\lambda$1397 $\approx 0.5$, \ion{O}{3}]
$\lambda$1663/\ion{He}{2} $\lambda$1640 $\approx 0.5$, and \ion{Si}{3}]
$\lambda$1892/\ion{C}{3}] $\lambda$1909 $\approx 0.3$. The emissivity
(or luminosity) weighted radius ($R_L$) is proportional to the emission
line response function centroid that in turn is equal to the continuum
-- emission line cross-correlation function centroid for linear line
responses to continuum variations (Koratkar \& Gaskell 1991; GOG93).
For this and all satisfactory solutions to the time-averaged spectrum
we find the following sequence in increasing $R_L$: \ion{N}{5},
\ion{Si}{4}, \ion{He}{2}, \ion{C}{4}, Ly$\alpha$, \ion{C}{3}],
\ion{Mg}{2}. This order very nearly corresponds to the one of
increasing emission line lags observed in NGC~5548. Thus a simple LOC
model can reproduce the observed spectrum and the general observed
trends in ionization stratification within the BLR of this and other
objects.

\subsection{Emission Line Light Curves and Lags from an LOC Model}

Using the emission line emissivities, the above adopted model
parameters ($R_{out}$, $\Gamma$, $C_f$), and the simple geometrical
assumptions ($\S$~2.2.2) we computed one dimensional emission line
response functions, $\Psi(\tau) \propto \eta(R)R^2F(R)$, where $F(R)$
is the emission line surface flux at radius $R$, $\eta(R)$ is the
responsivity of the cloud at radius $R$, and the lag $\tau =
\frac{R}{c}(1 + \rm{cos}~\theta\/$), with $\theta$ measuring the
azimuthal angle. We then convolved these emission line response
functions with observed UV $\lambda$1350 continuum light curves (C91;
K95) to generate the emission line light curves and emission line --
continuum cross correlation function (CCF) peak and centroid (at 50\%
peak) lags (Peterson \& White 1994). We present these light curves and
lags here, and discuss their comparisons with the observations in the
next section.

Figure~5a shows the comparison between the simulated (emissivity and
responsivity-weighted) and observed light curves for Ly$\alpha$,
\ion{N}{5}, \ion{Si}{4}, \ion{C}{4}, \ion{He}{2}, \ion{C}{3}], and
\ion{Mg}{2} (recall that several of these lines are blends) from the
1988--1989 {\em AGN Watch} {\em IUE} campaign for NGC~5548 (C91).
Figure~5b shows a similar comparison for the same sets of lines minus
\ion{Mg}{2} from the 1993 {\em HST} campaign (K95). In the latter we
also utilized a smoothed version of the measured noisy UV continuum
measured by {\em IUE} just prior to the {\em HST campaign}. The error
bars on the observed data points do not reflect the systematic errors
present to varying degrees in these data (C91; K95). The model
emissivity and responsivity-weighted CCF peak and centroid lags are
given in columns 4--7 of Table~2; the measured lags are reported in C91
and K95 and we discuss these further in the next section. The
responsivity $\eta(R)$ of an emission line is proportional to the slope
$dF_{line}/dF_{cont}$ (GOG93), and is a function of radius in our
simple spherical BLR. We used the ``local'' responsivity approximation,
in that at every radius each emission line was assigned the local value
of the response to a small variation in the continuum flux, given the
luminosity/flux normalization for the time-averaged spectrum. This is
appropriate as long as either the responsivity does not change
dramatically with radius or the continuum variations are not too
large.  While the first assumption does break down at small radii (see
Figure~3), these clouds generally do not contribute substantially to
the integrated emission line luminosities. Note that we have
approximated what should be $\eta(R, n_H, N_H)$ as $\eta(R)$. Finally,
when generating the simulated emission line light curves, we did not
alter the shape of the continuum. The opt-UV continuum in NGC~5548 has
been demonstrated to harden with increasing luminosity of the continuum
source (e.g., Romano \& Peterson 2000). Marshall et al.\ (1997) showed
that over short time intervals at least the EUV continuum is correlated
with but varies with a larger amplitude than does the UV continuum.
However, the detailed nature of continuum variability across the energy
bands remains a mystery (Nandra et al.\ 1998). Kaspi \& Netzer tried a
variety of different schemes to alter the SED with the UV luminosity in
order to produce a better match to the observed light curves. They
found that if their adopted SED's EUV break-point energy shifted from 3
Rydbergs to 5 Rydbergs with increasing UV luminosity, their models
could better reproduce the \ion{He}{2} light curve by increasing this
line's variability amplitude. The variability amplitude of Ly$\alpha$
increased as well, which was an improvement, although its mean flux
value became too large, and the overall quality of their fits to
Ly$\alpha$, \ion{C}{4}, \ion{C}{3}, and \ion{Mg}{2} diminished.

We emphasize that {\em at no point did we attempt to fit the observed
light-curves or lags} --- we made an {\em ad~hoc} choice of parameters
($R_{out}$, $\Gamma$, $C_f$) from a range of solutions which produced a
match to the integrated mean emission line spectrum within the
uncertainties, and which would result in a reasonable spread in the
emission line lags. This was done purposely to test whether or not
broad but simple cloud distribution functions which lead to matches to
a time-averaged spectrum might also predict the continuum -- line
reverberation. Additionally, given our present simplistic approach to
the LOC picture, we saw no reason to over-fit the data.

\section{DISCUSSION OF RESULTS AND THE FUTURE}

\subsection{Lags}

The emission lines' luminosity-weighted radii in Table~2 can be roughly
eye-balled in Figure~2 by mentally centroiding the EW contours
(allowing for the integrand limits in gas density and radius), since
for $\Gamma = \beta = -1$ the EW contours are directly proportional to
those of luminosity. This was pointed out by Baldwin et al.\ (1995).
However, the emission line lags will be biased toward the response of
emission line gas from smaller radii which can respond more rapidly and
more coherently to the continuum flux variations than gas at larger
radii. This explains in part why the predicted emissivity-weighted lags
are 3 to 5 times smaller than the corresponding values of $R_L/c$
(P\'{e}rez, Robinson, \& de la Fuente 1992). The model
responsivity-weighted lags will be generally longer than the
emissivity-weighted lags because the responsivity $\eta(R)$ is
proportional to the slope $dF_{line}/dF_{cont}$ which generally
flattens at small radii for most lines due to effects of ionization and
thermalization. 

Since the CCF, used to measure the emission line lags, is equivalent to
the convolution of the emission line response function with the
autocorrelation function of the continuum, the measured lags for an
emission line will depend upon the variability nature of the driving
continuum, even if the parameters which govern the distribution of
emitting gas in phase space are time steady. A continuum variation with
a characteristic time scale $\tau\/_{cont}$ will most effectively probe
line-emitting regions at distances $R \sim c\tau_{cont}/(1 +
\rm{cos}~\theta\/)$. Differences in the emission line lags are observed
between the two campaigns (C91; K95) and predicted in Table~2. These
differences may also occur due to a line's luminosity-weighted radius
that migrates in and out with the mean ionizing luminosity of the
continuum source (O'Brien, Goad, \& Gondhalekar 1995). This is a
consequence of non-linearity in the emission line response, and is
accommodated to some degree in our locally-linear response
approximation. Two other possible reasons for the observed changes in
the emission line lags are: (1) the BLR is non-stationary on a time
scale of 4 years that separates the campaigns (Wanders \& Peterson
1996), and (2) the finite nature of the monitoring campaigns coupled
with the dominance of long time-scale trends in the continuum flux
variations (Welsh 1999).

The predicted lags of Ly$\alpha$ and \ion{C}{4} in Table~2 lie fairly
near their reported values for the two monitoring campaigns (12 days
and 8 -- 16 days respectively [C91]; 7.5 -- 6.9 days and 4.6 -- 7.0
days, respectively [K95]). Those of the subordinate lines \ion{N}{5}
and \ion{He}{2} are long compared to their observed values (4 days and
4 -- 10 days, respectively [C91]; 1.4 -- 2.4 days and 1.7 -- 1.8 days,
respectively [K95]), while those of the $\lambda$1900 blend and
\ion{Mg}{2} are too short compared to their observed values (26 -- 32
days and $\ga 34$ days, respectively [C91]).  The predicted lags of the
$\lambda$1400 blend are too short compared to the measured values from
1989 campaign spectra ($\ga 12$ days [C91]), and perhaps a bit too long
compared to the values measured from the 1993 campaign spectra (3.5 --
4.8 days [K95]). We compare the model and observed light curves in
Figures~5a,b, and we will discuss these in more detail in $\S$~3.2.
Assuming these differences to be significant, they suggest some clues
as to how the actual gas distribution may differ from the one we have
derived from the mean spectrum, and we speculate here.

That the observations suggest longer \ion{Mg}{2} lags relative to that
of Ly$\alpha$ than produced in this model may imply the presence of
high density gas at radii larger than our model's outer radius (refer
to Figure~2b). Or perhaps the high density gas at larger radii
intercepts a larger fraction of the incident continuum than our
monotonic radial covering fraction function would predict. This
component may be denser than $10^{12}$~cm$^{-3}$ and may also be in
part the same gas which emits the Balmer emission lines (see K97) not
modeled here. We also suggest a possible reason that Kaspi \& Netzer's
models generally far underproduced \ion{Mg}{2} emission: at the larger
radii where this line is emissive, their pressure law resulted in
clouds with gas densities that emit \ion{Mg}{2} less than optimally.
Their imposed outer radius of 100 light-days also did not help in this
respect. On the other hand, it is our models' inclusion of this gas
that emits \ion{Mg}{2} $\lambda$2800 and little else other than Balmer
emission that helps drive the predicted integrated cloud covering
fractions to $\ga 40\%$. It must also be kept in mind that the
conditions under which \ion{Mg}{2} $\lambda$2800 is emitted are
affected by the radiative transfer of the Balmer lines and continuum,
and so is probably the least accurate of the seven line and line blends
simulated here.

In most of the models computed here, the lag of the $\lambda$1900 blend
is predicted to be just a bit longer than that of Ly$\alpha$, whereas
the observations show significantly longer lags for the $\lambda$1900
blend, albeit with considerable uncertainties. A glance at Figure~2a
shows that the $\lambda$1900 blend has a secondary peak in optimal
emission near the coordinates ($\log n_H \approx 9.25, \log \Phi_H
\approx 19$). This emission is almost entirely that of \ion{C}{3}]
$\lambda$1909 (K97) and lies at a far higher ionization parameter than
the main diagonal ridge of optimal emission ($\log U_H \approx -2.5$)
for \ion{C}{3}] and the blend. Lying near the $10^{23}$~cm$^{-2}$
column density-imposed ionization ``cliff,'' this emission's strength
depends much more sensitively to column density --- it forms near the
back boundary of the cloud for column densities $N_H \ga
10^{22}$~cm$^{-2}$. Tests show that an integration over a
$10^{22}$~cm$^{-2}$ column density grid with identical boundary
conditions to those here result in a 10\% increase in the
luminosity-weighted radius of $\lambda$1900 relative to that of
Ly$\alpha$. The presence of a range of column densities, such that the
lower gas density clouds (that are emissive in opt-UV lines only at
larger radii) have predominantly lower column densities, would further
separate the model lags of the $\lambda$1900 and Ly$\alpha$ lines. All
else being equal this would also reduce somewhat the predicted
$\lambda$1900 intensity as well as the \ion{C}{3}]/\ion{Si}{3} emission
line ratio.

Finally, it is seen in Figure~2a that the \ion{He}{2} blend emission
peaks near the $10^{23}$~cm$^{-2}$ column density-imposed ionization
``cliff'' for gas densities $9.5 \la \log n_H \la 11.5$. Clouds at
smaller radii than this blend's peak in Figure~2a would emit more
efficiently in both \ion{He}{2} $\lambda$1640 and \ion{O}{3}]
$\lambda$1663 if these clouds had higher column densities. Tests show
that an integration over a $10^{24}$~cm$^{-2}$ column density grid with
identical boundary conditions to those here result in a 15\% decrease
in the luminosity-weighted radius of the \ion{He}{2} blend relative to
that of Ly$\alpha$. The presence of a range of column densities, such
that higher column densities were more prevalent for gas densities $\ga
10^{11}$~cm$^{-3}$, would further separate the model lags of the
\ion{He}{2} blend and Ly$\alpha$. To a smaller extent this would also
apply to the \ion{N}{5} line as well. These adjustments may not be
enough, however, to overcome the significant disparities between the
predicted and observed \ion{He}{2} blend and \ion{N}{5} emission line
lags with respect to the continuum. The various problems concerning the
measured intensities and lags of the \ion{He}{2} spectrum in Seyfert~1
galaxies are to be discussed in greater detail in a work in progress
(Ferland et al.\ 2000).

\subsection{Light Curves}

The predicted emission line lags tell only part of the story. As
pointed out by Kaspi \& Netzer, the light curves of many emission lines
contain far more constraints than either just a mean spectrum or the
mean spectrum plus emission line lags. The offsets between the model
and observed light curves are for the most part explained by the
differences between the model and observed mean broad emission line
luminosities (Tables~1 and 2). That the model light curves match as
well as they do those from a campaign that occurred 4 years prior to
the {\em HST} campaign is remarkable (Figure~5a). This may mean that
the ``cloud'' parameters are reasonably steady on this time scale (but
see Wanders \& Peterson 1996). The responsivity-weighted emission line
light curves generally came closer to matching the observed light
curves. We invite the reader to compare the results in Figure~5a to
those of the more restrictive single pressure law models of Kaspi \&
Netzer (1999; Figures~7 and 9) for the five emission lines in common.
It should be kept in mind that in the latter work, the uni-dimensional
radial power-law cloud parameters were optimized to fit the 5 emission
line light curves explicitly.

Disregarding the simple offsets between the model and observed mean
luminosities, the greatest mismatches in Figure~5 occur in the
variation amplitudes of the two recombination lines, Ly$\alpha$ and
\ion{He}{2} (blended with \ion{O}{3}]), for both campaigns. The model
light curves of Kaspi \& Netzer (1999) for these two lines suffered
similarly until they allowed for a continuum luminosity dependent SED,
as mentioned above. Because of this deficiency the observed inverse
correlation between the continuum flux and the \ion{C}{4}/Ly$\alpha$
ratio (Pogge \& Peterson 1992) is not predicted by the model presented
here. A variable SED and/or the inclusion of a range of cloud column
densities (Shields, Ferland, \& Peterson 1995), as discussed above,
may resolve this shortcoming.

\subsection{Future Directions}

The most important feature of the LOC model is its assumption of a
large pool of ``line emitting entities'' from which to draw the
emission, and the strong influence of the natural selection effects
introduced by the atomic physics and general radiative transfer. While
reasonably successful in result, the analysis presented here is
actually quite restrictive in its approach to the LOC model. Here and
in Baldwin et al.\ (1995) the cloud parameter distributions were
limited to simple power law functions, and the gas density distribution
function was fixed over the full breadth of the broad line region. We
have also considered only a single column density. Nature need not
choose this rigid distribution, and indeed we have discussed how the
loosening of some of these assumptions might improve the model's match
to the observations. We do not know yet the origin of the ``line
emitting entities'' and the LOC model does not directly address this
origin, except that it does not need to be one of finely tuned
parameters. The LOC model requires only that there be a wide range of
``cloud'' properties throughout the BEL geometry. We point out that
while the LOC model was developed within the cloud paradigm, other
origins for the line emitting matter may fall within its general
philosophy. For example, an illuminated wind from an accretion disk
(e.g., Murray \& Chiang 1998), while differing in detail, has an
illuminating continuum cutting through gradients in gas density and
column density within some differential spherical radius. Such
gradients will depend upon the angle of the emitting gas above the disk
midplane.

The strength in the approach adopted here is in its simplicity in
reproducing general emission line properties of both composite quasar
spectra (Baldwin et al.\ 1995) and those of a well observed AGN.
However, the derivation of the detailed properties of the broad line
emitting regions and thus understanding their true nature will require
loosening the simplifying restrictions imposed here. In fact one could
hope to {\em derive} a more general multi-dimensional cloud
distribution function whose form need not be analytic, e.g., $f(n_H,
N_H, R, \theta\/, v_R)$ using time-variable spectra of sufficient
quality, where $\theta$ is the azimuthal angle and $v_R$ is the radial
velocity. Using fake AGN integrated emission line flux and continuum
flux light curves and a two-dimensional distribution function $f(n_H,
R)$, Horne, Korista, \& Goad (1999) outlined a means to unify the
methods of maximum entropy echo-mapping (Horne, Welsh, \& Peterson
1991; Krolik et al.\ 1991) with photoionization modeling. In this
approach, a general cloud distribution function is constrained mainly
by the data and the multi-dimensional spectral simulation grid, rather
than partly by the simplifying analytic assumptions made here. Horne et
al.\ also take into account the effects of anisotropic line emission
for simple cloud geometries, and consider a range of symmetric BLR
geometries. This new technique combines the direct and indirect methods
for solving this complex problem. In later work we will apply this
method to the light curves of the 1989 {\em IUE} monitoring campaign of
NGC~5548, with the hope of learning something concrete about the
distribution of the line emitting entities within its broad line
region. Eventually we hope to incorporate a more general cloud
distribution function, such as $f(n_H, N_H, R, \theta\/, v_R)$, in our
analysis. However, to take full advantage of this we may need a data
set of higher quality (mainly longer duration) than even the 1993 {\em
HST} campaign. These analyses should impose boundary conditions upon
the distribution functions describing the BEL gas, and therefore
constrain scenarios for the physical origins and dynamics of this gas.

\section{CONCLUSIONS}

Spanning over 10~billion years of cosmic history and 5~orders of
magnitude in energy, the general similarity of quasar/AGN spectra is
astounding. Baldwin et al.\ (1995) suggested the locally
optimally-emitting clouds (LOC) picture as a path to advancing our
understanding the broad emission lines of AGN: that selection effects
(atomic physics and general radiative transfer) operating in a large
``pool'' of line-emitting environments govern the spectra of quasars.
In this picture, the physical characteristics of the line-emitting
entities (e.g. gas density, column density) are not unique but are
broadly distributed along the radial dimensions of the BLR. The
continuum SED and gas chemical abundances are the primary drivers of
this natural selection process. In the case of NGC~5548 we find that
the ionizing continuum SED must be significantly harder than that of
Mathews \& Ferland (1987) to reproduce the observed
\ion{C}{4}/Ly$\alpha$ emission line ratio --- in concurrence with
interpolation of contemporaneous multi-wavelength observations, and the
gas abundances are roughly solar with some hint that some elements have
sub-solar abundances. More accurate derivations of each of these will
require analyses beyond the scope of this paper.

Using very simple (power law) ``cloud'' distribution functions in both
radius from the central source and gas density for a fixed cloud column
density, we tested the LOC picture against the spectroscopic
observations of NGC~5548. For a fixed but broad cloud distribution
function in gas density, the outer radius ($R_{out}$), power law index
of the radial cloud covering fraction function ($\Gamma$), and the
integrated cloud covering fraction ($C_f$) were optimized to predict
the time-averaged UV spectrum from the 1993 {\em HST} campaign.
Satisfactory fits to the time-averaged emission line spectrum are
possible for a broad range of parameters: $R_{out} \ga 60$ light-days,
$-1.6 \la \Gamma \la -0.5$, and $0.33 \la C_f \la 0.80$, although the
condition that $C_f \la 50\%$ (as well as \ion{O}{6}/\ion{C}{4}~$\leq
0.5$) constrains $\Gamma \ga -1.4$. The maximum outer radius of the BLR
is only loosely constrained by a time and velocity-averaged emission
line spectrum, but considerations of grain survival at low incident
continuum photon fluxes, the breadths of the emission line profiles,
and the observed responses of the emission lines probably conspire to
limit $R_{out} \la 200$ light-days. Consideration of the individual
emission line light curves would have placed stronger constraints upon
$R_{out}$, $\Gamma$ and $C_f$, had we chosen to do so.

Given that we did not optimize our models to fit directly the light
curves of the emission lines, but merely that of the time-averaged
spectrum from the 1993 {\em HST} campaign, the similarities of the
predicted emission line light curves and their lags compared to the
observed ones are remarkable. We believe this is a demonstration of the
natural predictive power of the LOC picture. The differences between
the observed emission line light curves and lags and those predicted by
our model suggest to us the following possible general improvements to
the LOC model presented here. A range of column densities is present
such that higher column density clouds are predominant for higher gas
density clouds, while lower column density clouds are predominant for
the lower gas density clouds. Because taken together most opt-UV
emission lines are visible only for a range in $\bar{U}_H \propto
\bar{L}/(R^2n_H)$ up to some maximum value of $\bar{U}_H$, a broadly
characteristic radial dependence of column density may be involved.
While this is broadly consistent with the findings of Kaspi \& Netzer,
there is no reason to believe that such column densities should be
uniquely defined at every radius. The gas density distribution function
need not remain constant over all radii as assumed here, and in
particular there may be a greater concentration of very high density
gas at the larger radii than our simple cloud distribution functions
would allow.

The next step involves loosening the simplifying constraints imposed
here and deriving a general distribution function of cloud properties,
e.g., $f(n_H, N_H, R, \theta\/, v_R)$, constrained mainly by the observed
spectra and the spectral simulations. This can be realized through the
marriage of echo-mapping techniques with spectral simulation grids,
using the constraints provided by a high quality temporal spectroscopic
data set.

Finally, we wish to respond to a statement that appears in Kaspi
\& Netzer (1999). They wrote ``Finally, we must comment that present
day LOC models are too general and do not contain full treatment of
shielding and mixing of the various coexisting components.'' We are not
certain what was meant by ``are too general.'' Being ``general'' is the
main point of the LOC picture. We have shown here and elsewhere the LOC
picture has been applied, that there is a fairly wide range in the way
that BLR gas can be distributed in ($R, n_H, N_H$) space and still
produce spectra that match typical quasar/AGN spectra. Here we fit the
mean integrated fluxes of a series of lines in a specific AGN. Many
LOC/pressure-law models can do that, but to match the variability
requires that the mean formation radius of the lines show a large
degree of variation and in such a way that high ionization lines form
characteristically closer in to the central ionizing continuum source.
As first pointed out by Baldwin et al., this appears to be a natural
consequence of the LOC picture. By contrast, in the single pressure-law
model, a given pressure-law constrains the run of density and column
density with radius to vary in a very specific manner, and their
starting values must be normalized to a given ionization
parameter/incident flux at a pre-defined radius. Amongst other things
the LOC picture's generality frees one to investigate some important
global parameters over the population of quasars, such as the continuum
SED and gas abundances (e.g., Korista, Baldwin, and Ferland 1998). In
regards to the ``shielding and mixing'' comment, the models of broad
emission line spectra of AGN presented here, in Baldwin et al.\ (1995),
and elsewhere are and have been internally self-consistent. We
discussed a more mature approach to the LOC picture in $\S$~3.3.

\acknowledgements

We thank an anonymous referee for his/her constructive comments. This
work benefited substantially from the support of a PPARC grant of Keith
Horne's and we would like to thank Keith and the University of
St.\ Andrews for their hospitality. MRG acknowledges support through a
PPARC fellowship during the completion of this work. We are also
grateful to Gary Ferland for maintaining his freely distributed code,
Cloudy, and to Jack Baldwin for his inspiration. We also thank Jack for
his careful reading and suggestions which improved the manuscript.

%
%

%
\clearpage

\begin{figure}
\figurenum{1}
\caption{The observed mean UV FOS/{\em HST} spectrum of the
Seyfert~1 galaxy, NGC~5548. The flux units are
$10^{-13}~\rm{ergs~s^{-1}~cm^{-2}}$~\AA\/$^{-1}$. The emission features
lying just shortward of the strong broad emission line of Ly$\alpha$
$\lambda$1216 belong to geocoronal Ly$\alpha$.}
\end{figure}

\begin{figure}
\figurenum{2a}
\caption{Contours of $\log W_{\lambda}$~(EW) for six emission lines (or
blends), referenced to the incident continuum at 1215~\AA\/ for full
source coverage, are shown as a function of the hydrogen density and
flux of hydrogen-ionizing photons. The total hydrogen column density is
$10^{23}~\rm{cm^{-2}}$. The EW is in direct proportion to the continuum
reprocessing efficiency. The smallest, generally outermost, decade
contour corresponds to 1~\AA\/, each solid line is 1 decade, and dotted
lines represent 0.1 decade steps. The contours generally decrease
monotonically from the peak to the 1~\AA\/ contour; the solid triangle
marks the location of the peak of the dominant line within the blends
discussed in the text (Ly$\alpha$, \ion{He}{2}, \ion{C}{3}], and
\ion{Si}{4}). The solid star is a reference point marking the old
``standard BLR'' parameters.}
\end{figure}

\begin{figure}
\figurenum{2b}
\caption{Same as Figure~2a for the emission line \ion{Mg}{2}
$\lambda$2800.}
\end{figure}

\begin{figure}
\figurenum{3}
\caption{The emission line/blend radial surface fluxes from the
model clouds for the mean ionizing luminosity, given the adopted
weighting function along the gas density axis. The radial distance is
measured from the continuum source and the vertical solid line marks 1
light-day.}
\end{figure}

\begin{figure}
\figurenum{4a}
\caption{Envelopes in minimum $\chi^2$ (solid curves) as
functions of each of the three parameters, as determined by the
simulated annealing process. The lower dashed lines show the $1\sigma$
confidence level ($\chi^2 = \chi^2_{min} + 1.00 = 2.01$) for one
interesting parameter. The upper dashed lines show the $1\sigma$
confidence contour ($\chi^2 = \chi^2_{min} + 4.72 = 5.73$) for $N - M =
7 - 3 = 4$ interesting parameters.}
\end{figure}

\begin{figure}
\figurenum{4b}
\caption{Confidence contours of $\log \chi^2$ as a function of $C_f$
and $\Gamma$ for fixed values of $\log R_{out}$ incremented at 0.2 dex.
The contours are in steps of 0.25 dex with the outer value equal to
1.50 dex in every case. Satisfactory models lie within the bold dashed
contour representing $\chi^2 = \chi^2_{min} + 4.72 = 5.73$.}
\end{figure}

\begin{figure}
\figurenum{5a}
\caption{Continuum and {\em broad} emission line light curves from the
1989 {\em IUE} monitoring campaign of NGC~5548. The solid points with
error bars are the observations, the bold curves are the model
emissivity weighted predictions and the lighter, generally lower
amplitude variation curves are the model responsivity weighted
predictions. The vertical axes are in luminosity units: $10^{40}$
$\rm{ergs~s^{-1}}$~\AA\/$^{-1}$ for the UV continuum and $10^{42}$
$\rm{ergs~s^{-1}}$ for the UV emission lines.}
\end{figure}

\begin{figure}
\figurenum{5b}
\caption{Continuum and {\em broad} emission line light curves from the
1993 {\em HST} monitoring campaign of NGC~5548. The solid points with
error bars are the observations, the bold curves are the model
emissivity weighted predictions and the lighter, generally lower
amplitude variation curves are the model responsivity weighted
predictions. The vertical axes are in luminosity units: $10^{40}$
$\rm{ergs~s^{-1}}$~\AA\/$^{-1}$ for the UV continuum and $10^{42}$
$\rm{ergs~s^{-1}}$ for the UV emission lines.}
\end{figure}

\clearpage

%
\begin{deluxetable}{lcccc}
\small
\tablecaption{Mean UV Emission Line Strengths in NGC~5548. \label{tbl-1}}
\tablewidth{0pt}
\tablehead{
\colhead{Emission Line or Blend} &
\multicolumn{1}{c}{Observed Total} & \multicolumn{1}{c}{Narrow Line} &
\multicolumn{1}{c}{Corrected BEL} & \multicolumn{1}{c}{$\log
\bar{L}_{BEL}$}\\
\colhead{} & \colhead{Flux} & \colhead{Flux} & \colhead{Flux} & \colhead{}
}
\startdata
Ly$\alpha$ $\lambda$1216 & 694 & 89.5 & 808 & 42.66 $\pm$ 10\% \\
\ion{N}{5} $\lambda$1240 & 79.4 & 6.6 & 96.4 & 41.74 $\pm$ 50\% \\
\ion{Si}{4} + \ion{O}{4}] $\lambda$1400 & 75.0 & 7.6:: & 85.6 & 41.69 $\pm$ 30\% \\
\ion{C}{4} $\lambda$1549 & 676 & 69.7 & 757 & 42.64 $\pm$ 10\% \\
\ion{He}{2} $\lambda$1640 + \ion{O}{3}] $\lambda$1663 & 99.1 & 6.0 & 116 & 41.82 $\pm$ 30\% \\
\ion{C}{3}] $\lambda$1909 + \ion{Si}{3}] $\lambda$1892 & 123 & 13.6 & 138 & 41.90 $\pm$ 30\% \\
\ion{Mg}{2} $\lambda$2800 & 129 & $\sim$~0 & 152 & 41.94 $\pm$ 50\% \\
\enddata

\tablecomments{Only the major components of the line blends are listed.
Fluxes are observed frame and given in units of
$10^{-14}~\rm{ergs~s^{-1}~cm^{-2}}$. Luminosities are given in units of
ergs~s$^{-1}$. All mean values taken from the 1993 {\em HST} campaign
(K95) with the exception of \ion{Mg}{2}. See text for further details.}

\end{deluxetable}

\clearpage

%
\begin{deluxetable}{lcccccc}
\small
\tablecaption{Model Time-Averaged \& Time-Dependent Properties of the UV
Spectrum. \label{tbl-2}}
\tablewidth{0pt}
\tablehead{
\colhead{Emission Line or Blend} & \colhead{$\log \bar{L}$} & 
\colhead{$R_L/c$} &
\multicolumn{2}{c}{IUE89} & \multicolumn{2}{c}{HST93}\\ 
\colhead{} & \colhead{} & \colhead{} & \colhead{$\tau_L$} & 
\colhead{$\tau_{\eta}$} & \colhead{$\tau_L$} & 
\colhead{$\tau_{\eta}$}
}
\startdata
Ly$\alpha$ $\lambda$1216 & 42.64 & 41.4 & 10--10.6 & 10--12.2 & 7--9.3 & 8--10.3 \\
\ion{N}{5} $\lambda$1240 & 41.85 & 12.8 & 3--4.7 & 5--5.4 & 3--5.6 & 4--6.6 \\
\ion{Si}{4} + \ion{O}{4}] $\lambda$1400 & 41.70 & 16.0 & 5--5.9 & 7--8.3 & 4--6.7 & 6--8.4 \\
\ion{C}{4} $\lambda$1549 & 42.68 & 29.2 & 7--8.8 & 10--11.5 & 6--8.3 & 7--10.2 \\
\ion{He}{2} $\lambda$1640 + \ion{O}{3}] $\lambda$1663 & 41.74 & 22.1 & 6--6.4 & 6--7.8 &
4--7.1 & 5--8.1 \\
\ion{C}{3}] $\lambda$1909 + \ion{Si}{3}] $\lambda$1892 & 41.82 & 42.3 & 11--13.0 & 14--16.6 & 8--10.7 & 13--12.8 \\
\ion{Mg}{2} $\lambda$2800 & 41.83 & 67.3 & 15--19.7 & 21--22.7 & 15--13.6 & 17--14.9 \\
\enddata

\tablecomments{Only the major components of the line blends are
listed. Luminosities are given in units of ergs~s$^{-1}$. The light
travel time to the luminosity-weighted radius, $R_L/c$, and the
emissivity and responsivity-weighted CCF lags, $\tau_L$ and
$\tau_{\eta}$, are given in units of days. Each pair of lags in columns
4 -- 7 is given as ``CCF peak -- CCF centroid,'' the latter measured at
50\% of the CCF peak.}

\end{deluxetable}

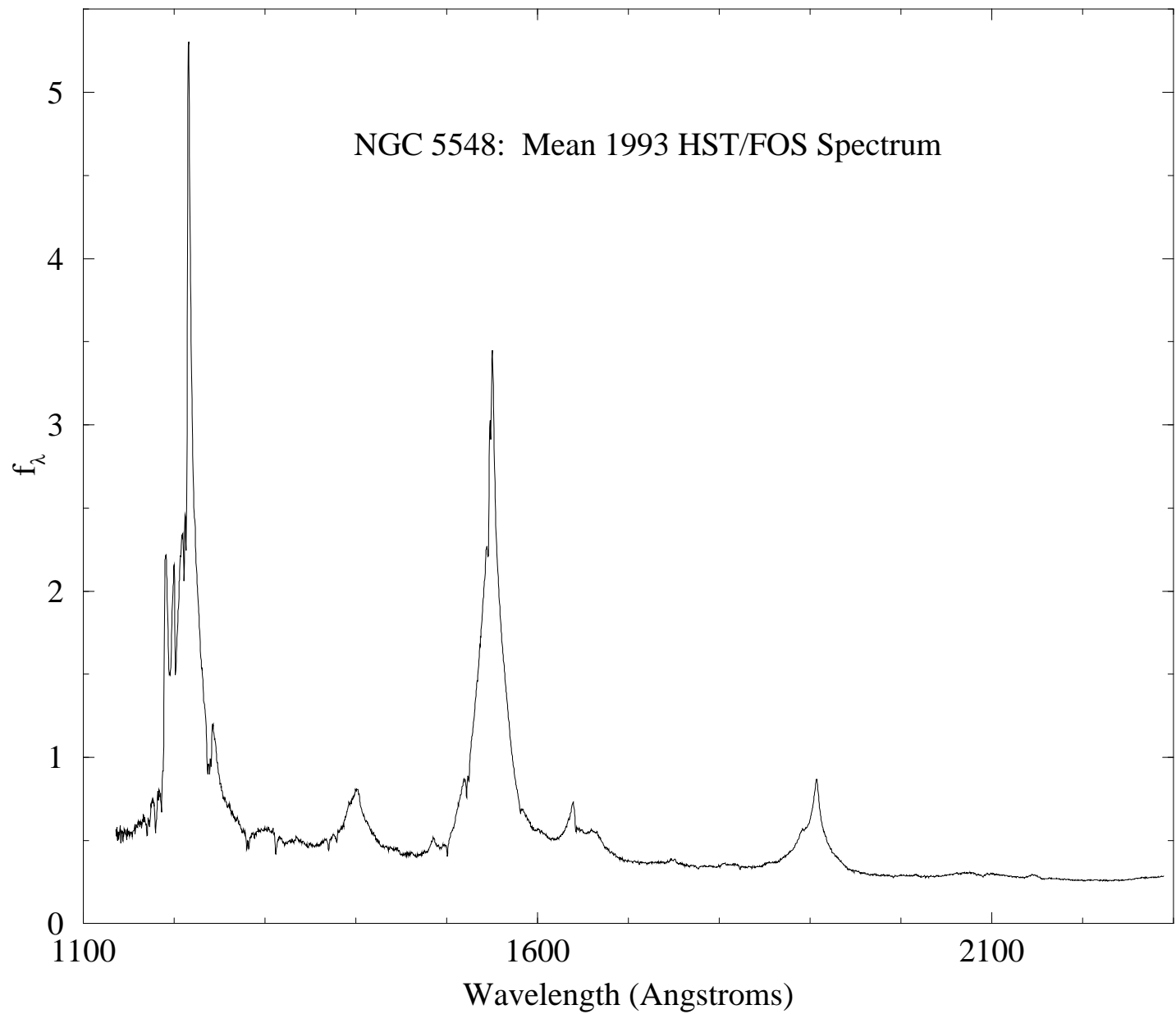
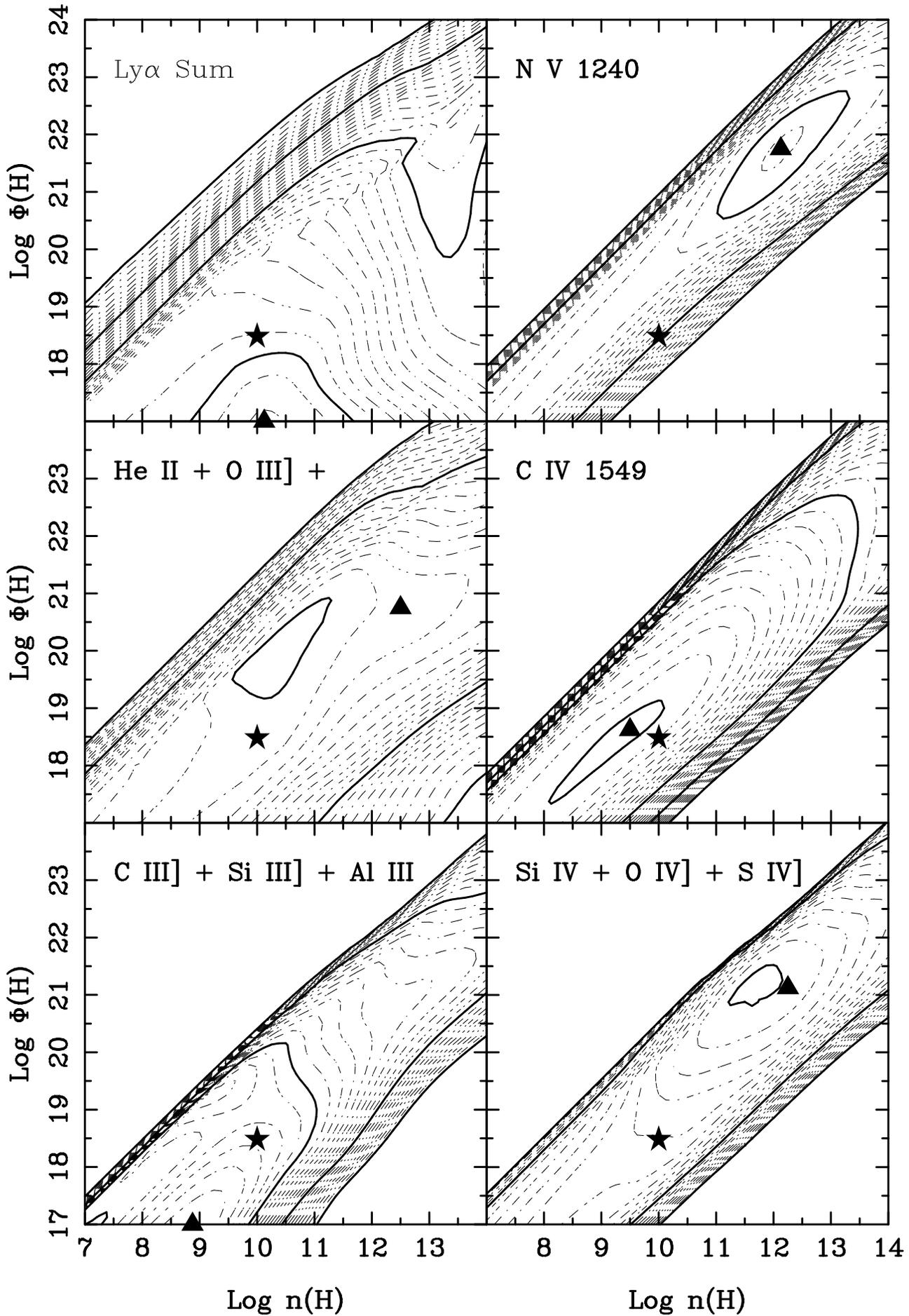
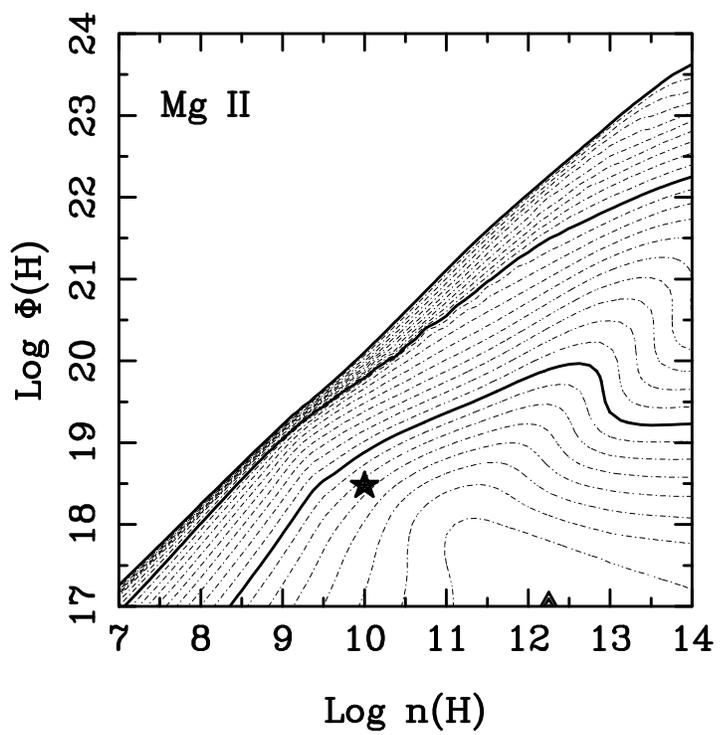
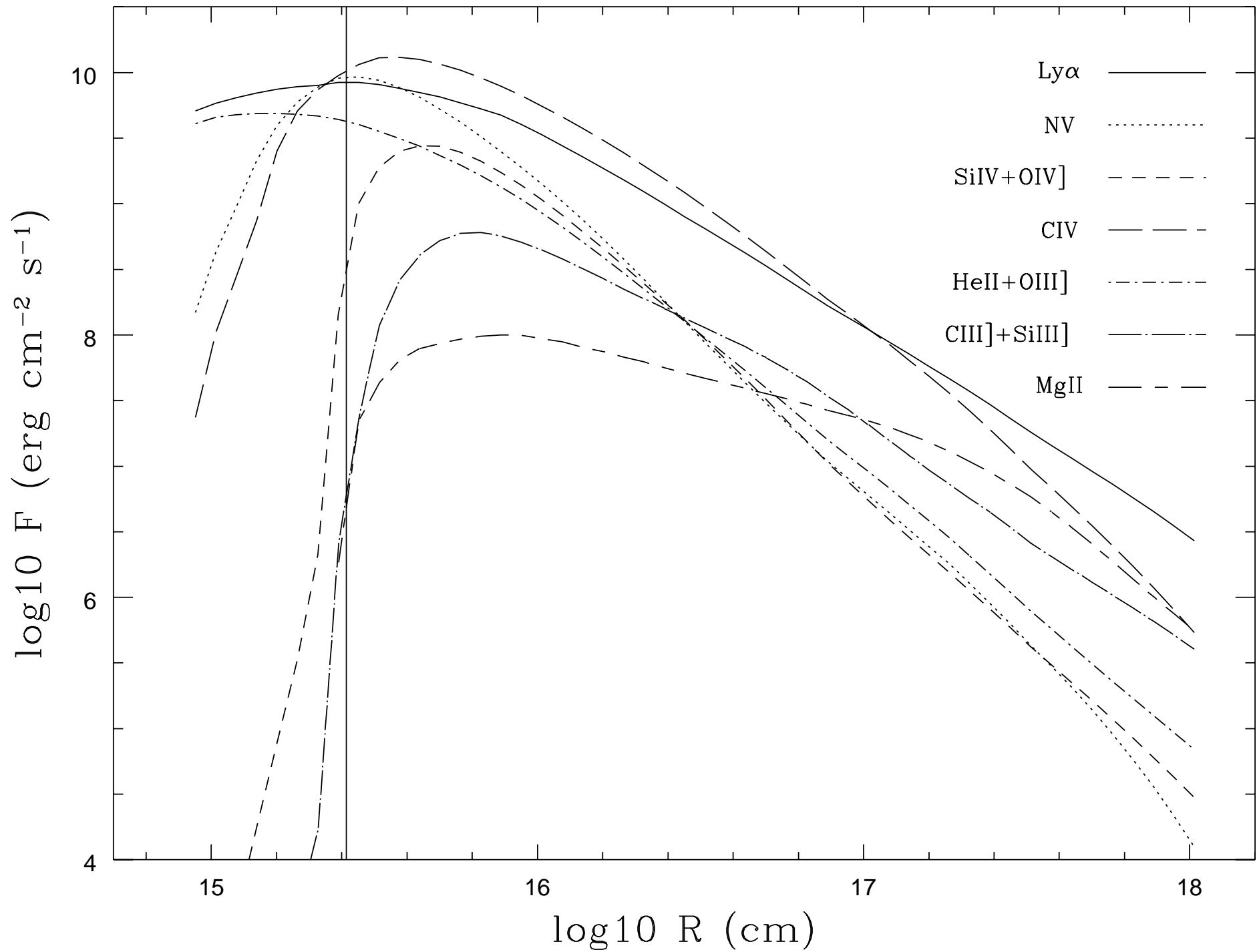
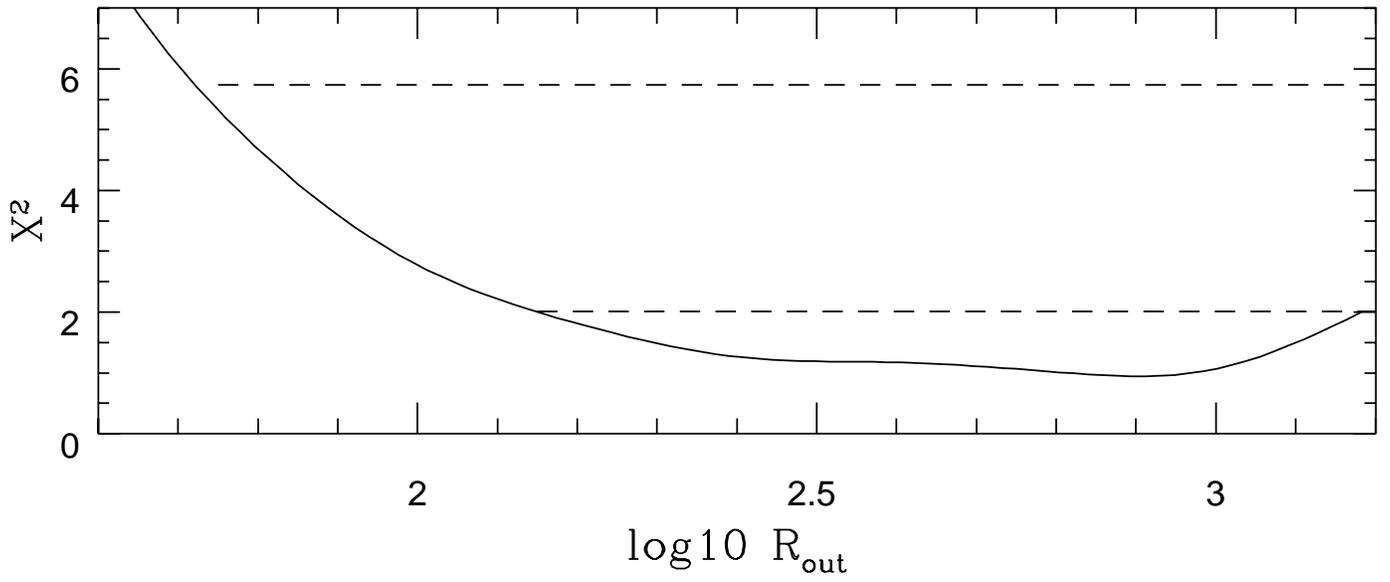
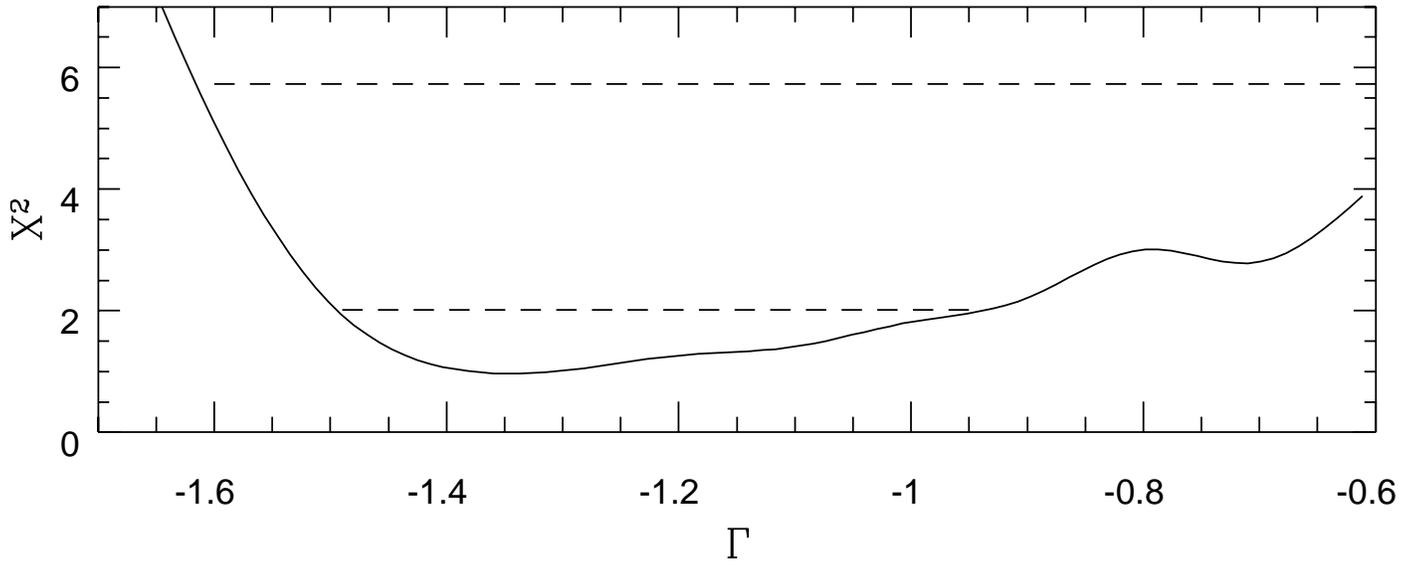
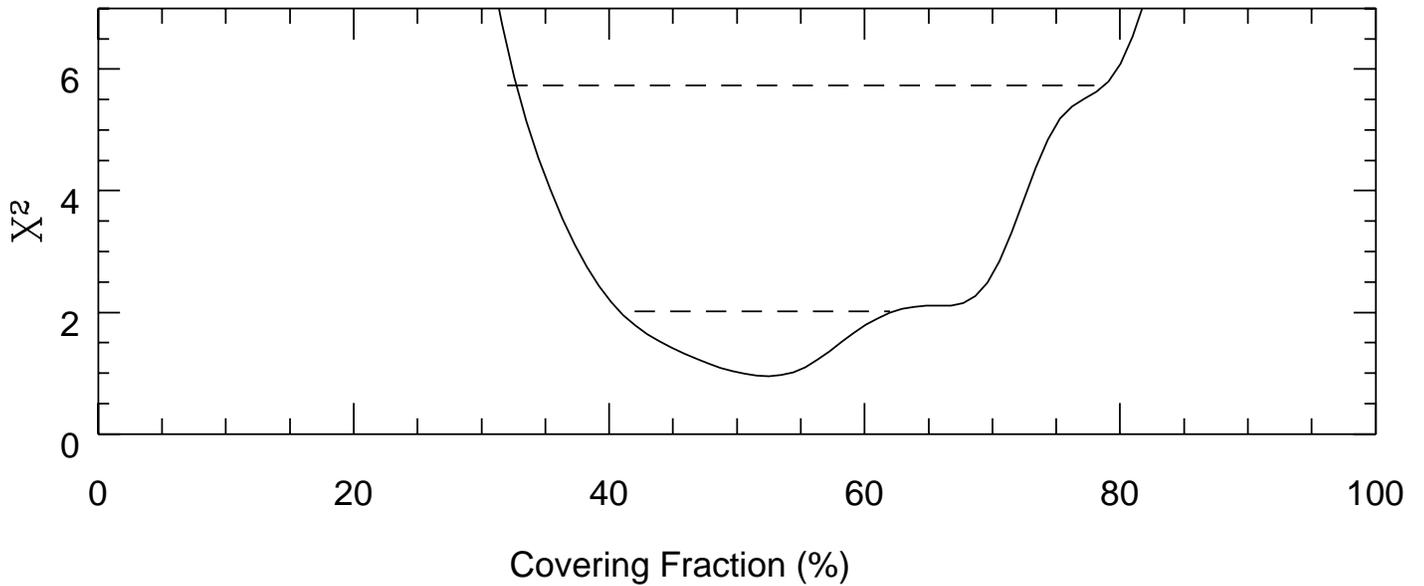
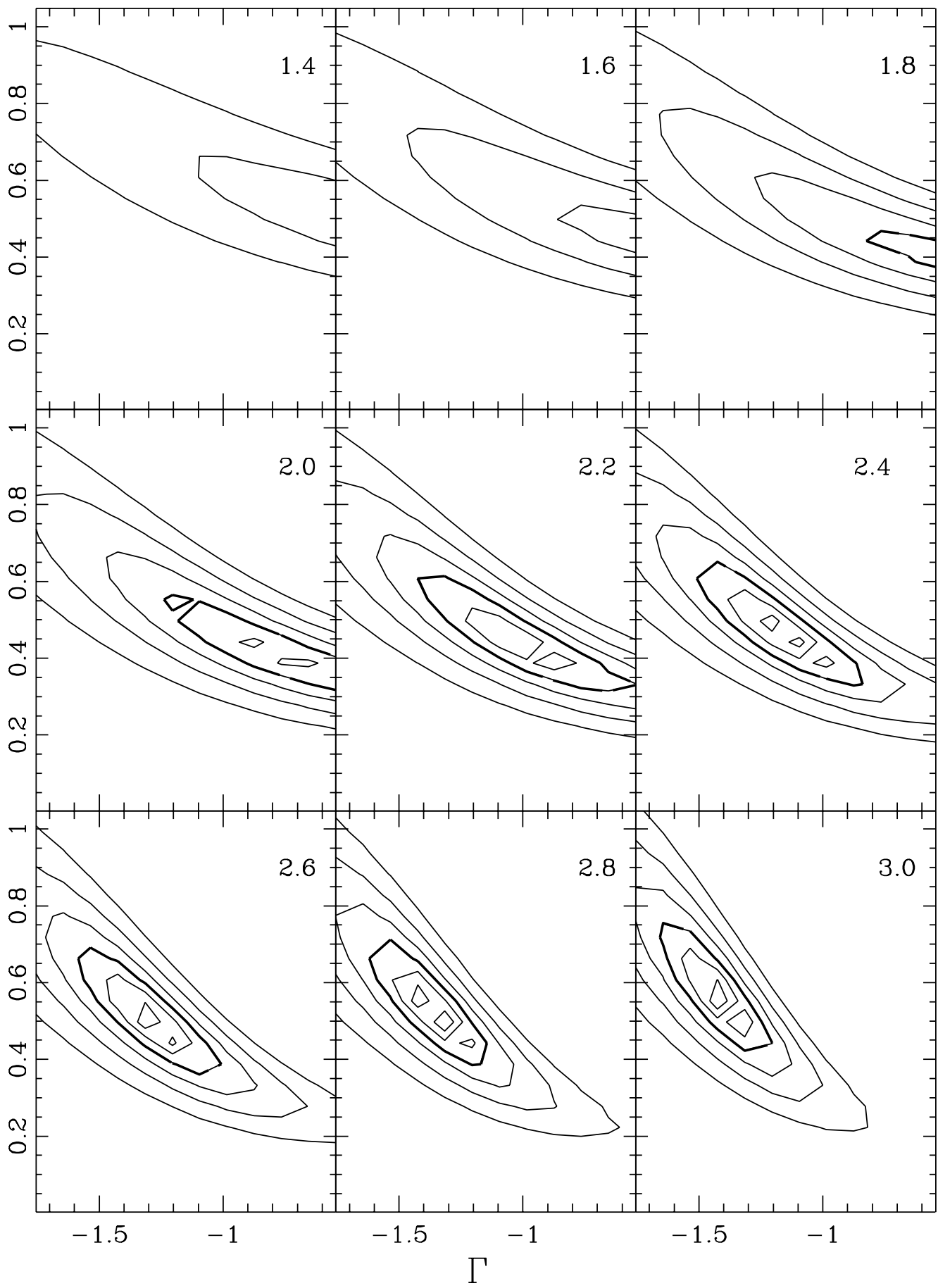
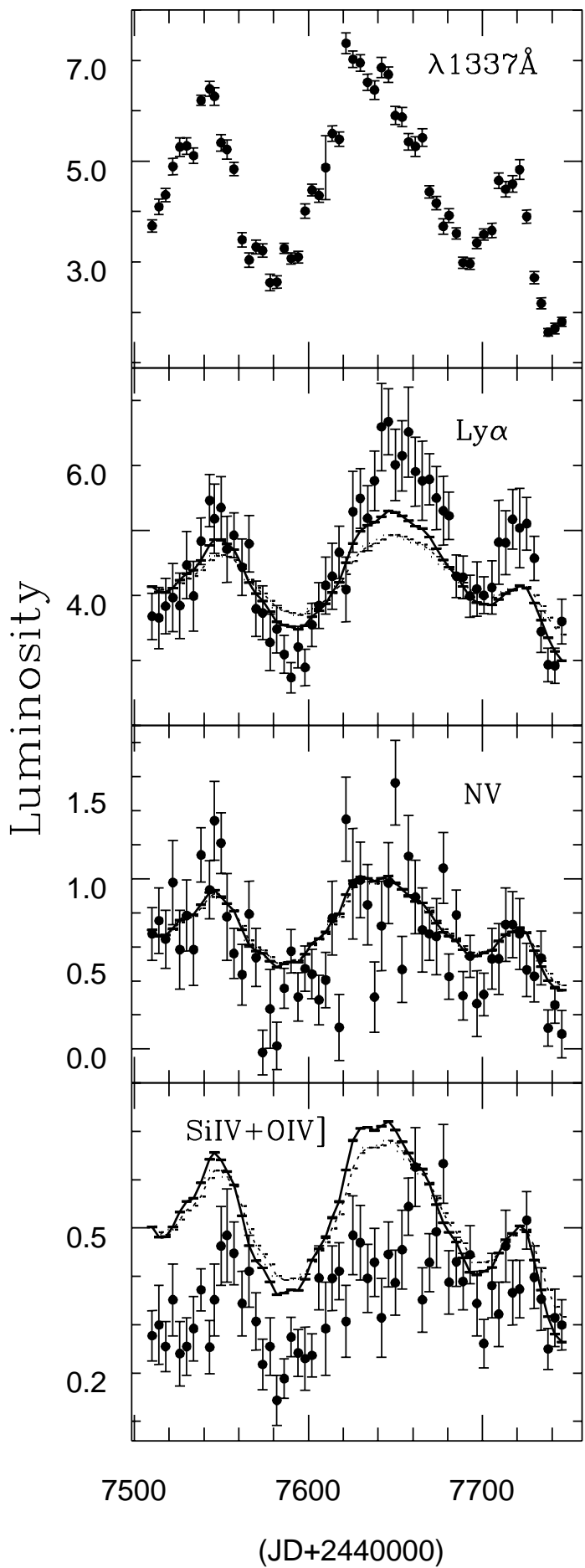
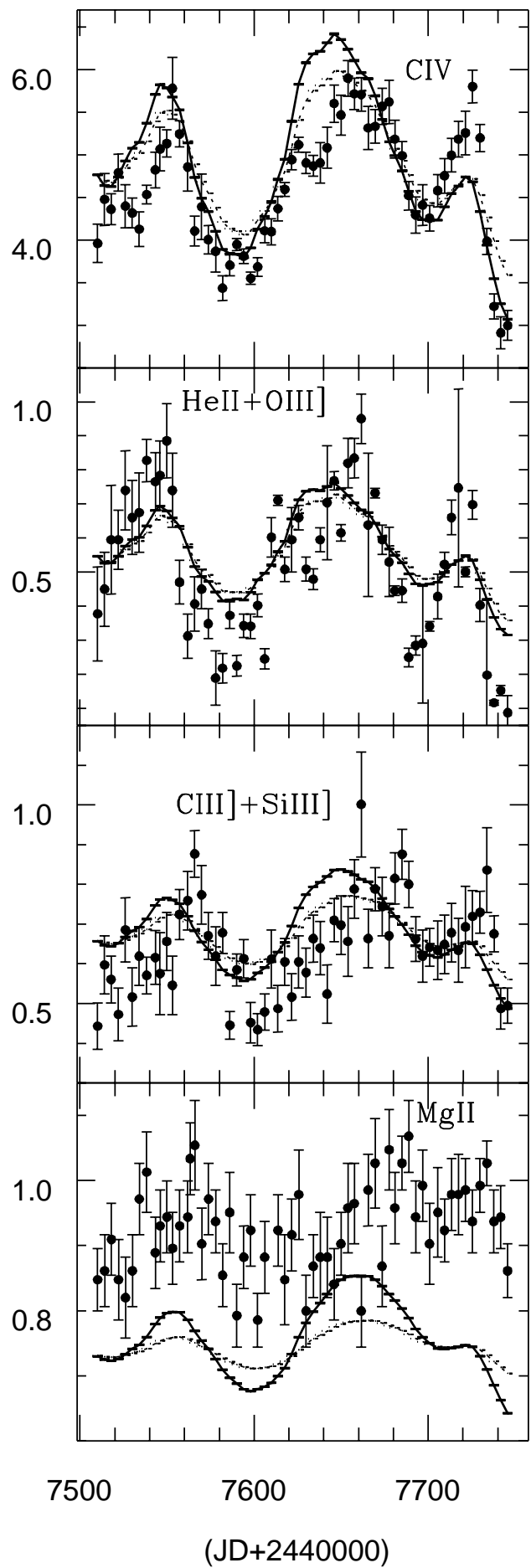
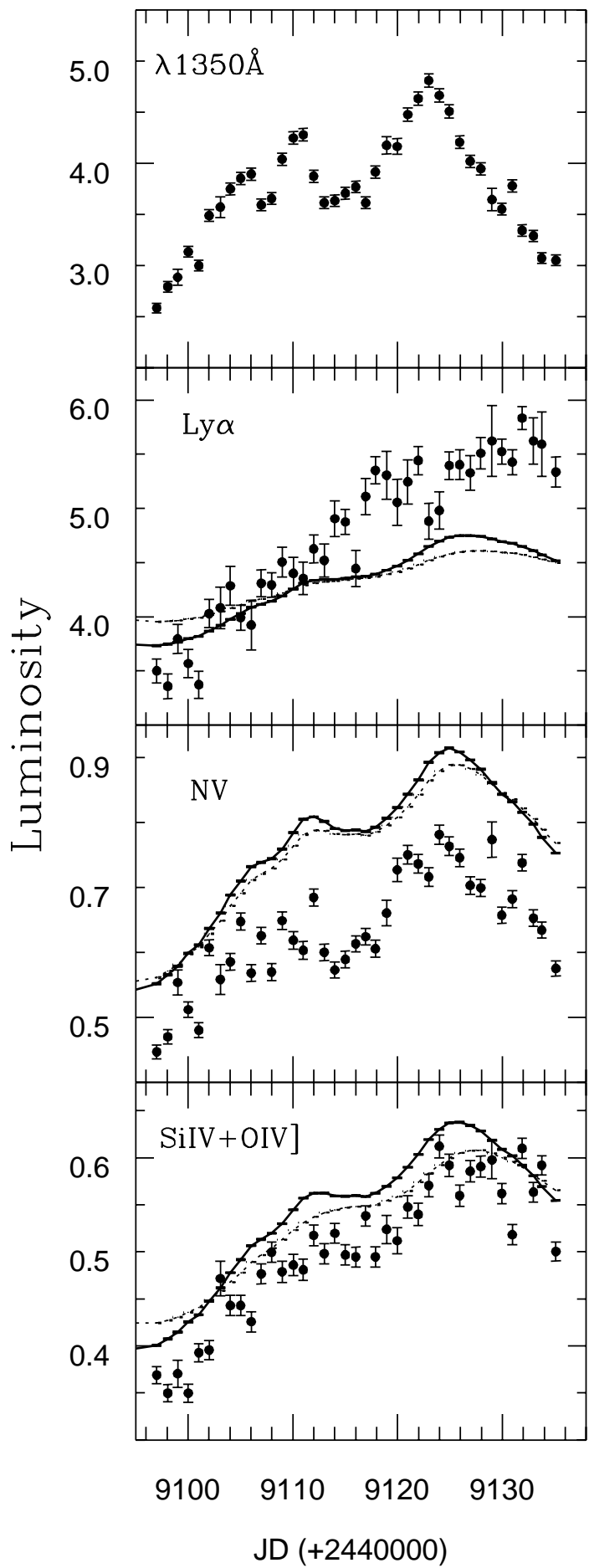
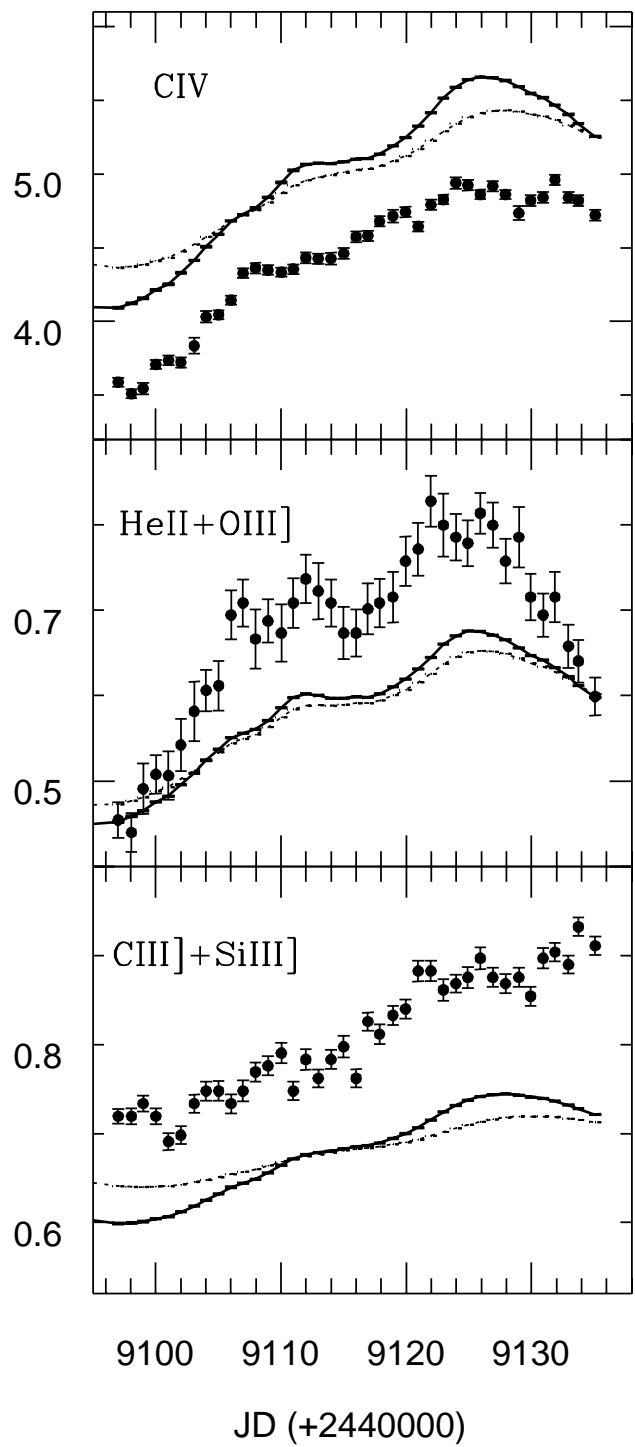
\end{document}